\documentclass[showpacs,twocolumn,prl,showkeys]{revtex4-1}
\usepackage[dvips]{graphicx}
\usepackage{amsmath,amssymb,amsthm,mathrsfs,amsfonts,dsfont}
\usepackage{subfigure, epsfig}
\usepackage{braket}
\usepackage{bm}
\usepackage{enumerate}
\usepackage{multirow}
\usepackage{color}
\usepackage{qcircuit}
\usepackage{lineno}

\usepackage{bm}
\renewcommand{\vec}[1]{{\ensuremath{\bm{\mathrm{#1}}}}}  
\usepackage{xspace}
\newcommand{\AFIC}{antiferromagnetic\xspace}
\newcommand{\AFM}{antiferromagnets\xspace}
\newcommand{\MnSn}{Mn$_{3}$Sn\xspace}
\newcommand{\MnGe}{Mn$_{3}$Ge\xspace}

\begin{document}

\title{Infrared Imaging of Magnetic Octupole Domains in Non-collinear Antiferromagnets}

\begin{abstract}
Magnetic structure plays a pivotal role in the functionality of \AFM (AFMs), which not only can be employed to encode digital data but also yields novel phenomena. Despite its growing significance, visualizing the \AFIC domain structure remains a challenge, particularly for non-collinear AFMs. Currently, the observation of magnetic domains in non-collinear antiferromagnetic materials is  feasible only in \MnSn, underscoring the limitations of existing techniques that necessitate distinct methods for in-plane and out-of-plane magnetic domain imaging. In this study, we present a versatile method for imaging the \AFIC domain structure in a series of non-collinear \AFIC materials by utilizing the anomalous Ettingshausen effect (AEE), which resolves both the magnetic octupole moments parallel and perpendicular to the sample surface. Temperature modulation due to the AEE originating from different magnetic domains is measured by the lock-in thermography, revealing distinct behaviors of octupole domains in different antiferromagnets. This work delivers an efficient technique for the visualization of magnetic domains in non-collinear AFMs, which enables comprehensive study of the magnetization process at the microscopic level and paves the way for potential advancements in applications.
\end{abstract}


\author{Peng Wang$^{1,2,\dagger}$}
\author{Wei Xia$^{3,4,\dagger}$}
\author{Jinhui Shen$^{1,5}$}
\author{Yulong Chen$^{1,5}$}
\author{Wenzhi Peng$^{1,5}$}
\author{Jiachen Zhang$^{1,5}$}
\author{Haolin Pan$^{1,5}$}
\author{Xuhao Yu$^{1,5}$}
\author{Zheng Liu$^{5,6}$}
\author{Yang Gao$^{5,6}$}
\author{Qian Niu$^{5,6}$}
\author{Zhian Xu$^{3}$}
\author{Hongtao Yang$^{7}$}
\author{Yanfeng Guo$^{3,4,*}$}
\author{Dazhi Hou$^{1,5,*}$}

\affiliation{$^1$ICQD, School of Emerging Technology, University of Science and Technology of China, Hefei 230026, China}
\affiliation{$^2$College of Mathematics and Physics, Qingdao University of Science and Technology, Qingdao 266061, China}
\affiliation{$^3$School of Physical Science and Technology, ShanghaiTech University, Shanghai 201210, China}
\affiliation{$^4$ShanghaiTech Laboratory for Topological Physics, Shanghai 201210, China}
\affiliation{$^5$Department of Physics, University of Science and Technology of China, Hefei, Anhui 230026, China}
\affiliation{$^6$CAS Key Laboratory of Strongly-Coupled Quantum Matter Physics, University of Science and Technology of China, Hefei, Anhui 230026, China}
\affiliation{$^7$Xi'an Institute of Optics and Precision Mechanics of Chinese Academy of Sciences, Xi'an, Shanxi 710119, China}

\maketitle

\begin{flushleft}
	$^*$Corresponding authors: \\
	guoyf@shanghaitech.edu.cn;\\
	dazhi@ustc.edu.cn.\\
	$^\dag$These two authors contributed equally to this work.\\
\end{flushleft}

~\clearpage
\section{INTRODUCTION}\label{sec1}

The non-collinear antiferromagnet {Mn$_{3}$X\xspace} family has emerged as a significant class of materials in the field of spintronics\cite{Higo2022,Tsai2020,Liu2018,Qin2023,Chen2023,Jeon2023,Deng2022}. These materials possess unique physical properties, including the magnetic spin Hall effect\cite{Hu2022}, the large anomalous Hall/Nernst effects (AHE/ANE)\cite{Nakatsuji2015,Ikhlas2017,Li2017-ANE}, the Weyl semimetal points\cite{Kuroda2017}, and the chiral domain wall memory effect\cite{Li2019}. Recent progress further demonstrated their electrical manipulation and readout capabilities, making them promising building elements for future memory devices\cite{Qin2023,Chen2023}.

Therefore, the domain structure visualization in noncollinear antiferromagnet (AFM) is greatly pursued\cite{Higo2018large,Reichlova2019,Uchimura2022,Yan2022,Sthr1999,Scholl2000,Hedrich2021,Krizek2022,Xu2019}, which is indispensable for confirming and understanding the magnetic octupole dynamics driven by magnetic field and spin orbit torque\cite{Takeuchi2021,Safeer2015,Tang2020,Liu2021,Wang2022,Xu2022}. Although the magneto-optical Kerr effect (MOKE) has been successfully used to image domain structures in {Mn$_{3}$Sn}\cite{Higo2018large,Uchimura2022}, it requires a pristine mirror surface and can image magnetic octupole domains with out-of-plane polarization with only few-nanometer depth\cite{Wu2020}. The anomalous Nernst effect, on the other hand, can image octupole domain with in-plane polarization, but is limited to in-plane observations and is time-consuming for large devices due to the scanning approach\cite{Reichlova2019}. Despite the early interests, the direct imaging and reconstruction of 3D rotation of octupole moments, which calls for simultaneous observation of both in-plane and out-of-plane octupole domains, remains elusive.

\section{RESULTS AND DISCUSSION}\label{sec2}

\begin{figure*}
\centerline{\includegraphics[width=400pt]{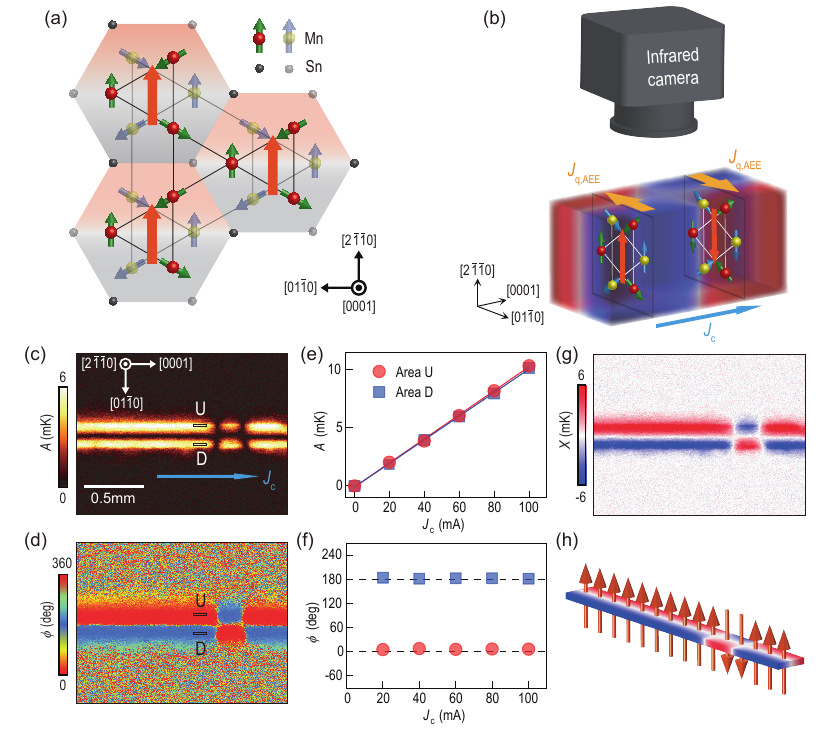}}
\caption{Magnetic structure of \MnSn and infrared imaging of magnetic domains by the AEE.
                \textbf{a}, Red and yellow spheres denote Mn atoms on two stacked kagome planes, and the arrows on the atoms represent the moments. The spin texture can be viewed as a ferroic order of a cluster magnetic octupole. The orange arrows represent the direction of the octupole moment.
                \textbf{b}, A schematic of infrared imaging of magnetic octupole domains with opposite polarization using the AEE. $\vec{J}_{\rm{c}}$ and $\vec{J}_{\rm{q,AEE}}$ denote the applied charge current and heat current generated by the AEE, respectively. 
			\textbf{c},\textbf{d},  Lock-in amplitude $A$ and phase $\phi$ images for the $(2\bar{1}\bar{1}0)$ plane of a \MnSn slab with an out-of-plane easy axis at $J_{\rm{c}}$ = 50 mA along the $[0001]$ direction.
			\textbf{e},\textbf{f}, $J_{\rm{c}}$ dependence of $A$ and $\phi$ in the regions of interest on the \MnSn slab which are defined by two rectangles marked with U and D. The plotted data was obtained by averaging the values in each area, which consists of 28 pixels.
			\textbf{g}, Image of $X = A\rm{cos}\phi$.
			\textbf{h}, Magnetic octupole domain patterns of the \MnSn slab revealed by the temperature modulation in \textbf{g}.
\label{fig1}}
\end{figure*}

Here, we demonstrate an alternative method for imaging the domain structure in non-collinear AFMs. We show that the domain structure can be imaged by the anomalous Ettingshausen effect (AEE) at room temperature, capturing octupole moments both parallel and perpendicular to the sample surface. The visualization of magnetic domains through the AEE was first demonstrated in ferromagnetic metals\cite{Uchida2018-AEE,Wang2020-LIT}. The method is based on visualizing the spatial distribution of AEE-induced temperature modulations originating from different magnetic octupole domains, and is achieved through lock-in thermography (LIT)\cite{breitenstein2010lock,Seki2018,Daimon2016}. Employing the new techique, we achieve the observation of the octupole domain structure in \MnSn and \MnGe during both in-plane and out-of-plane magnetic reversals and reveal the out-of-plane rotation process of octupole moments in the memory effect of Mn$_{3}$Sn\cite{Li2019}.

Figure 1a shows the magnetic structure of \MnSn, which hosts a hexagonal crystal structure with $P6_3/mmc$ space group. Below its N\'eel temperature $T_{\mathrm{N}} \approx$ 430 K, it exhibits an inverse triangular spin structure of three neighboring Mn moments on its (0001)-plane kagome lattice. This spin structure can be viewed as a ferroic order of a cluster magnetic octupole\cite{Suzuki2017}, which is comprised of six Mn atoms situated within two stacked kagome planes. The cluster magnetic octupole breaks the time-reversal symmetry macroscopically, which permits a non-zero net Berry curvature in momentum space\cite{Chen2014,Yang2017}, thereby eliciting significant transverse responses, such as the large AHE and ANE\cite{Nakatsuji2015,Ikhlas2017,Li2017-ANE,Xu2020,Chen2021Anomalous}. Additionally, the competition between the Dzyaloshinskii-Moriya interaction and single-ion anisotropy results in a small net ferromagnetic moment through spin canting\cite{Nagamiya1982}, which is essential for manipulating the direction of octupole moments using a magnetic field.

\begin{figure*}
\centerline{\includegraphics[width=400pt]{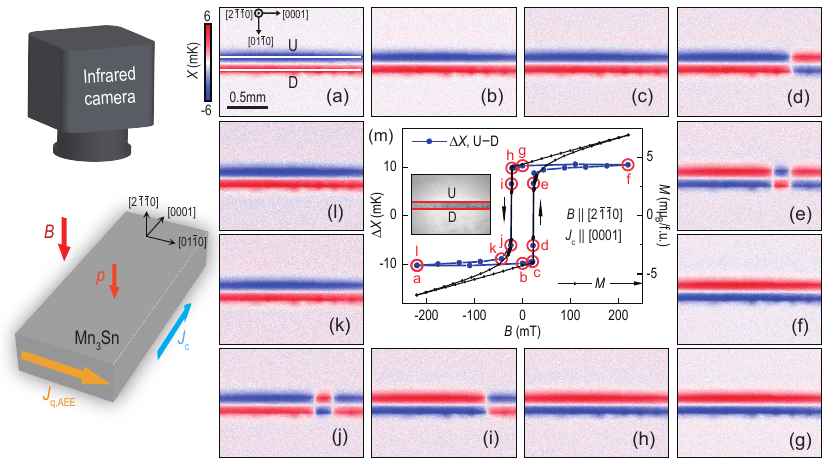}}
\caption{Infrared imaging of magnetic domain patterns in \MnSn during out-of-plane magnetic reversal.
                \textbf{a}-\textbf{l}, $X$ images obtained in the out-of-plane field-scan cycle from $-$220 mT to 220 mT along $[2\bar{1}\bar{1}0]$ at $J_{\rm{c}}$ = 50 mA along $[0001]$.
                \textbf{m}, Field dependence of $\Delta X$ calculated from the U and D areas of the \MnSn slab, where the plotted data were obtained by subtracting the average values on D from U, which consists of 640 pixels. For comparison, the field dependence of the magnetization $M$ is shown. Inset shows the DC infrared image of the $(2\bar{1}\bar{1}0)$ surface plane of the \MnSn slab. The thickness and width of the slab are approximately 0.09 mm and 0.17 mm, with a charge current density of $j_{\rm{c}} = 3.3 \times 10^{3}$ A m$^{-2}$.
\label{fig2}}
\end{figure*}

\subsection{Magnetic domain imaging by the AEE}
Figure 1b shows the set-up for infrared imaging of magnetic octupole domain structures using the AEE. In view of the large ANE in Mn$_{3}$Sn\cite{Ikhlas2017,Li2017-ANE}, a large AEE in \MnSn is expected by the Bridgman relation\cite{Bridgman1924} between AEE and ANE. The AEE can convert a longitudinal electric current $\vec{J}_{\rm{c}}$ into a transverse heat current $\vec{J}_{\rm{q,AEE}}$ that is dependent on the direction of the octupole moment in \MnSn. The heat current generates a thermal gradient $\nabla T_{\rm{AEE}}$, so the phenomena of AEE can be described as
\begin{equation}
\nabla T_{\rm{AEE}} = \varepsilon_{\rm{AEE}} (\vec{j}_{\rm{c}}\times\vec{p}),
\label{eq1}
\end{equation}
where $\vec{j}_{\rm{c}}$ is the charge current density, $\vec{p}$ is the unit vector of the octupole moment, $\varepsilon_{\rm{AEE}}$ is the anomalous Ettingshausen coefficient. Therefore, by measuring the AEE-induced temperature distribution on the sample surface, one can reveal the underlying octupole domain structures. 

In particular, when an in-plane current is applied, the domains with opposite out-of-plane $\vec{p}$ generate opposite in-plane thermal gradients on the sample surface (top view of Fig. 1b), while the domains with opposite in-plane $\vec{p}$ produce either a cold or hot region on the sample surface (front view of Fig. 1b). To capture both scenarios with a much-improved temperature resolution (0.1 mK), LIT is utilized\cite{breitenstein2010lock,Seki2018,Daimon2016}. In this work, thermal images of the sample surface are captured while applying a rectangular-wave-modulated charge current with amplitude $J_{\rm{c}}$, and frequency $f=$ 16.25 Hz. The use of such current allows for the extraction of thermoelectric effects ($\propto J_{\rm{c}}$) free from the Joule-heating background ($\propto J_{\rm{c}}^2$)\cite{Miura2019,Seki2019}. Fourier analyses are performed to extract the first harmonic response of the detected thermal images, yielding lock-in amplitude $A$ and phase $\phi$ images, with the $A$ images providing the magnitude of the AEE-induced temperature modulation and the $\phi$ images providing information on the sign of the temperature modulation and the time delay due to thermal diffusion.

Figures 1c and 1d, respectively, display the $A$ and $\phi$ images for a \MnSn slab with $(2\bar{1}\bar{1}0)$ oriented surface at zero magnetic field after oscillation demagnetization and $J_{\rm{c}}=$ 50 mA along the $[0001]$ direction. All \MnSn crystal samples used in this study were grown by bismuth flux method without polishing, from which Kerr signal can not be detected probably due to surface oxidization and roughness (see Supplementary Fig. S1). We observe three clearly distinct temperature-modulation regions, where the left and right regions show temperature increase and decrease in the up and down halves while the middle region shows an opposite pattern, indicating opposite in-plane temperature gradients along $[01\bar{1}0]$. For quantitative analysis, two regions of interest along $[01\bar{1}0]$, denoted as U and D, are defined on the \MnSn slab. In Figs. 1e and 1f, we show the $J_{\rm{c}}$ dependence of $A$ and $\phi$ for U and D. The plotted data is obtained by averaging the values in each area. The $A$ values are proportional to $J_{\rm{c}}$, while the $\phi$ values remain unchanged with respect to $J_{\rm{c}}$ and exhibit a 180-degree shift for the U and D regions, which is in good agreement with the features of the AEE with the local $\vec{p}$ along the $[2\bar{1}\bar{1}0]$ direction.

Figure 1g shows the $X=A\cos\phi$ image calculated from 1c and 1d. The use of $X=A\cos\phi$ to represent temperature modulation with sign information is more intuitive and easier to calculate the orientation of octupole domains in the absence of a significant phase delay. Therefore, throughout the following text, we adopt the $X$ image to present the spatial distribution of temperature modulation. Taking into account the magnetic anisotropy that tends to align $\vec{p}$ along the $[2\bar{1}\bar{1}0]$ direction\cite{Duan2015,Li2021easyaxis,Li2018easyaxis}, these observations strongly suggest that the sample has a domain distribution with upward oriented $\vec{p}$ on the left and right sides and downward oriented $\vec{p}$ in the middle. Assuming the AEE as the origin of the thermal patterns, we plot in Fig. 1h the magnetic octupole domain patterns of the \MnSn slab that are revealed by the $X$ image in Fig. 1g.

\begin{figure*}
\centerline{\includegraphics[width=400pt]{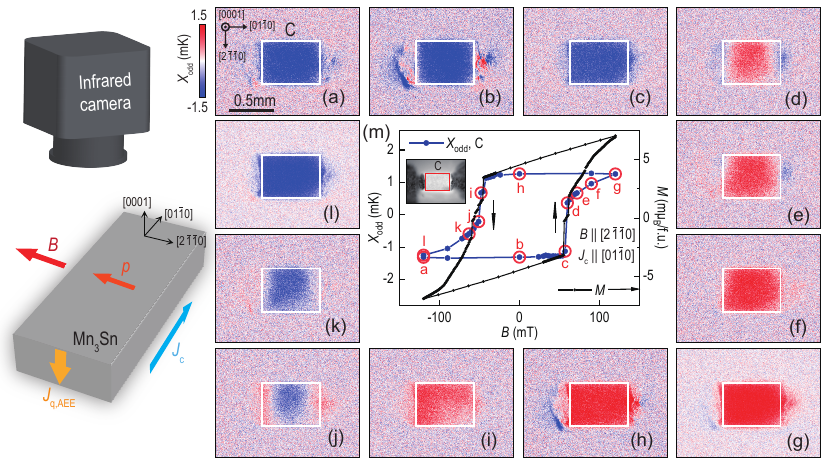}}
\caption{Infrared imaging of magnetic domain patterns in \MnSn during in-plane magnetic reversal.
                \textbf{a}-\textbf{l}, $X_{\rm{odd}}$ images obtained in the in-plane field-scan cycle from $-$120 mT to 120 mT along $[2\bar{1}\bar{1}0]$ at $J_{\rm{c}}$ = 50 mA along $[01\bar{1}0]$, where $X_{\rm{odd}}$ denotes the lock-in temperature modulation with the $B$-odd dependence.
                \textbf{m}, Field dependence of $X_{\rm{odd}}$ on the C area of the \MnSn slab, where the plotted data was obtained by averaging the values on C, which consists of 10800 pixels. For comparison, the field dependence of the magnetization $M$ is shown. Inset shows the DC infrared image of the \MnSn slab with the surface along the $(0001)$ direction. The thickness and width of the slab are approximately 0.12 mm and 0.45 mm, with a charge current density of $j_{\rm{c}} = 0.9 \times 10^{3}$ A m$^{-2}$.
\label{fig3}}
\end{figure*}

\subsection{Imaging the out-of-plane magnetic reversal of \MnSn}

To verify the validity of the AEE for imaging octupole domain structures, we applied it to visualize the domain reversal by external magnetic fields. Figure 2a–2l display a series of $X$ images of the same sample as in Fig.1c under the out-of-plane field-scan cycle from $-220$ mT to 220 mT along $[2\bar{1}\bar{1}0]$. The applied current $J_{\rm{c}}$ is 50 mA in the $[0001]$ direction, which generates a temperature gradient in the $[01\bar{1}0]$ direction (Fig. 2a). To evaluate the temperature difference along this direction, we defined two regions of interest, labeled as U and D (Fig. 2a, and inset of Fig. 2m), which respectively cover the upper and lower edges of the \MnSn slab. Fig. 2m plots the field dependence of $\Delta X$, which is calculated by subtracting the average $X$ values of D from those of U. The resulting $\Delta X$ curve exhibits a closed hysteresis loop, analogous to its AHE curve (see Supplementary Fig. S2a). For comparison, the field dependence of the magnetization $M$ is plotted in Fig. 2m. Apart from a linear background, $\Delta X$ curve is in good agreement with the $M$ hysteresis loop (see Supplementary Fig. S3a for further comparison). The $\Delta X$ curve obtained using this technique is similar to the temperature gradient results observed from transport measurements\cite{Xu2020}. However, the corresponding $X$ images provide visualization of the domain evolution at each point (i.e., points a-l) of the curve, which is beyond the capacity of transport measurements. Based on the calculated thermal gradient $\nabla T$ along $[01\bar{1}0]$, the AEE coefficient of \MnSn is estimated to be $\lvert \varepsilon_{\rm{AEE}} \rvert \approx 18$ $\mu$K m$\cdot$A$^{-1}$, which is the first experimentally determined value for this material. The  ANE coefficient $\lvert S_{\rm{ANE}} \rvert$ of \MnSn reported by Li \textit{et al}.\cite{Li2017-ANE} is about 0.5 $\mu$V K$^{-1}$ at room temperature, with a thermal conductivity $\kappa$ of approximately 7.4 W K$^{-1}$m$^{-1}$. Using the Bridgman relation $\varepsilon_{\rm{AEE}} = T  S_{\rm{ANE}} / \kappa$\cite{Bridgman1924}, this predicts an AEE coefficient $\lvert \varepsilon_{\rm{AEE}} \rvert$ of about 20 $\mu$K m$\cdot$A$^{-1}$, which is close to our measured value.

Figures 2b and 2g show the domain patterns under zero external magnetic field, which are nearly identical to those observed in Figs. 2a and 2f under saturation fields, confirming a spontaneous AEE at zero field whose sign can be switched by a coercive field $\sim$ 23 mT. Meanwhile, Figures 2d, 2e, 2i, and 2j reveal four intermediate states during the reversal process, showing three sizeable single-domain regions separated by two (0001) facets. These single-domain regions undergo a sharp reversal along the $[2\bar{1}\bar{1}0]$ easy axis even in a field scanning with field sweeps by a 0.01 mT step. It should be noted that although each domain region in this sample exhibits a sharp reversal along easy axis, the reversal behavior can vary significantly between samples (see Supplementary Fig. S4). Furthermore, the LIT imaging, as well as other imaging methods such as MOKE and ANE, can only visualize domain regions composed of a large number of magnetic octupoles and cannot resolve the spin structure of magnetic octupoles or the internal structure of domain walls.

\begin{figure*}
\centerline{\includegraphics[width=400pt]{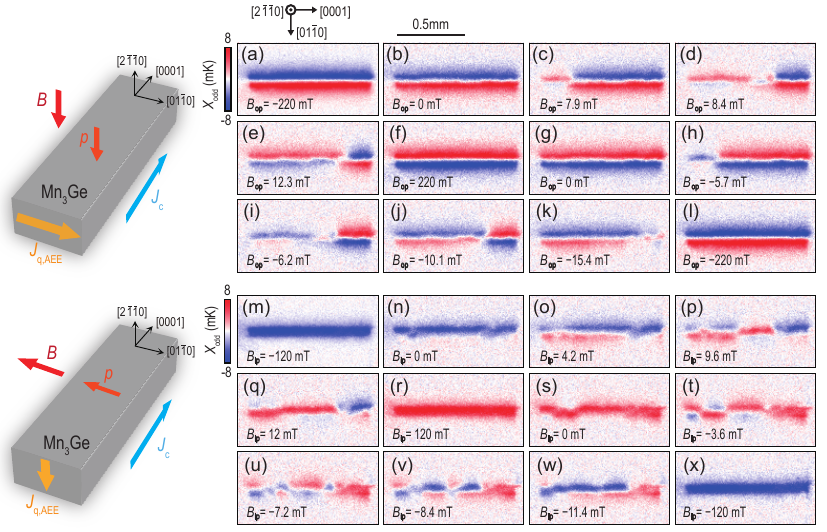}}
\caption{Infrared imaging of magnetic domain patterns in \MnGe during out-of-plane and in-plane magnetic reversals.
                \textbf{a}-\textbf{l}, $X_{\rm{odd}}$ images obtained in the out-of-plane field-scan cycle from $-$220 mT to 220 mT along $[2\bar{1}\bar{1}0]$ at $J_{\rm{c}}$ = 50 mA along $[0001]$.
                \textbf{m}-\textbf{x}, $X_{\rm{odd}}$ images obtained in the in-plane field-scan cycle from $-$120 mT to 120 mT along $[01\bar{1}0]$ at $J_{\rm{c}}$ = 50 mA along $[0001]$. $X_{\rm{odd}}$ denotes the lock-in temperature modulation with the $B$-odd dependence. 
\label{fig4}}
\end{figure*}

\subsection{Imaging the in-plane magnetic reversal of \MnSn}

Figure 3 presents the results of in-plane domain reversal of a \MnSn sample with the surface oriented along the $(0001)$ direction. Figure 3a-3l display the $X_{\rm{odd}}$ images obtained through an in-plane field-scan cycle ranging from $-$120 mT to 120 mT along $[2\bar{1}\bar{1}0]$ at $J_{\rm{c}}$ = 50 mA along $[01\bar{1}0]$. Here, $X_{\rm{odd}} = A_{\rm{odd}} \cos \phi_{\rm{odd}}$ refers to the lock-in temperature modulation with $B$-odd dependence, which eliminates a background primarily originating from the Peltier effect of the two contact electrodes\cite{breitenstein2010lock}, where $A_{\rm{odd}}e^{i\phi_{\rm{odd}}} = [Ae^{i\phi} - (A_{\rm{max}B}e^{i\phi_{\rm{max}B}}+A_{\rm{min}B}e^{i\phi_{\rm{min}B}})/2]$, and $A_{\rm{max}B}$ ($\phi_{\rm{max}B}$) and $A_{\rm{min}B}$ ($\phi_{\rm{min}B}$) represent the lock-in amplitude (phase) at the maximum positive magnetic field and minimum negative magnetic field, respectively. In this configuration, the temperature gradient of the AEE is in the out-of-plane direction, resulting in the formation of corresponding heating or cold cooling regions on the surface, depending on the direction of the in-plane octupole moment. Thus, the color change from blue to red and back to blue in Figs. 3a-3l corresponds to the in-plane reversal process along the $[2\bar{1}\bar{1}0]$ direction. Fig. 3m illustrates the field dependence of the averaged $X_{\rm{odd}}$ on the C region (defined in Fig. 3a, and inset of Fig. 3m), which exhibits a hysteresis loop. For comparison, the field dependence of the magnetization $M$ is also shown in Fig. 3m, and they are in a good agreement (Supplementary Fig. S3b). By combining the $X_{\rm{odd}}$ images at selected points on the hysteresis (e.g., Figs. 3d, 3e, 3j, and 3k), it is suggested that the in-plane reversal of this sample involves a rather gradual process of $\vec{p}$ rotation initially in the middle and then on both sides. The results in Fig. 2 and 3 demonstrate the effectiveness of the AEE as a versatile tool to observe both out-of-plane and in-plane octupole domain structures.

\subsection{Imaging the in-plane and out-of-plane magnetic reversals of \MnGe}

Figure 4 presents the results of out-of-plane and in-plane domain reversals of a \MnGe sample with the surface oriented along the $[2\bar{1}\bar{1}0]$ direction. Figure 4a-4l display the $X_{\rm{odd}}$ images obtained through an out-of-plane field-scan cycle ranging from $-$220 mT to 220 mT along $[2\bar{1}\bar{1}0]$ at $J_{\rm{c}}$ = 50 mA along $[0001]$. Analysis of Fig. 4c-4e and Fig. 4h-4j indicates that the out-of-plane domain reversal process in this sample proceeds from left to right, rather than being a sharp transition directly from up to down, passing through the $[01\bar{1}0]$ direction. Figure 4m-4x display the $X_{\rm{odd}}$ images obtained through an in-plane field-scan cycle ranging from $-$120 mT to 120 mT along $[01\bar{1}0]$ at $J_{\rm{c}}$ = 50 mA along $[0001]$. It is evident that the in-plane reversal process also occurs from left to right and involves traversing the $[2\bar{1}\bar{1}0]$ direction. The results presented in Fig. 4 exemplify the versatility of our infrared imaging technique in studying the {Mn$_{3}$X\xspace} family. The ANE coefficient $\lvert S_{\rm{ANE}} \rvert$ of \MnGe reported by Tomita \textit{et al}.\cite{Tomita2022-kappa} is about 0.35 $\mu$V K$^{-1}$ at 300 K, with a thermal conductivity $\mathrm{\kappa}$ of approximately 6.6 W K$^{-1}$m$^{-1}$. The AEE coefficient $\lvert \varepsilon_{\rm{AEE}} \rvert$ of \MnGe was first measured by Xu \textit{et al}.\cite{Xu2020}, which is about 15 $\mu$K m$\cdot$A$^{-1}$ at 300 K. The value we estimated by LIT is about 17 $\mu$K m$\cdot$A$^{-1}$, close to their value by transport measurements.

\begin{figure*}
\centerline{\includegraphics[width=400pt]{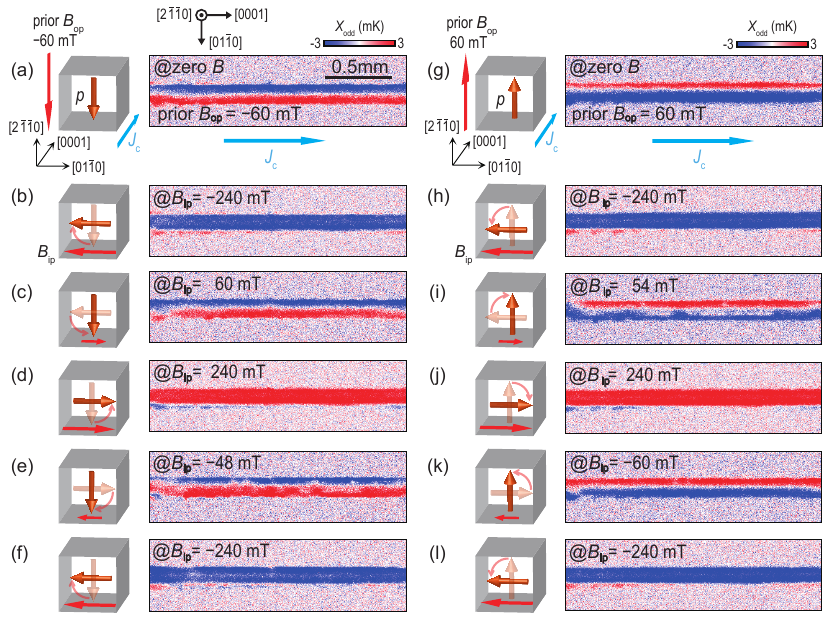}}
\caption{Visualization of the magnetic memory effect by the anomalous Ettingshausen effect in \MnSn.
                \textbf{a},\textbf{g}, The initial orientation of the octupole moment and the corresponding $X_{\rm{odd}}$ image obtained at zero field and $J_{\rm{c}}$ = 50 mA along $[0001]$, after applying a prior field $B_{\rm{op}}$ = $-$60 mT (\textbf{a}) and $B_{\rm{op}}$ = 60 mT (\textbf{g}) along $[2\bar{1}\bar{1}0]$. \textbf{b}-\textbf{f},\textbf{h}-\textbf{l}, The out-of-plane rotation of the octupole moment during an in-plane field-scan cycle from $-$240 mT to 240 mT along $[01\bar{1}0]$, and the corresponding measured $X_{\rm{odd}}$ images, with the prior field $B_{\rm{op}}$ = $-$60 mT (\textbf{b}-\textbf{f}) and $B_{\rm{op}}$ = 60 mT (\textbf{h}-\textbf{l}). The chirality of the octupole moment rotation is found to depend on the direction of the prior field.
\label{fig5}}
\end{figure*}

\subsection{Visualization of the magnetic memory effect}
Figure 5 presents the visualization for the prior-field-dependent chirality of octupole rotation during in-plane magnetic field sweep, which is due to the ``memory effect" in {Mn$_{3}$Sn}\cite{Li2019}. Fig. 5a illustrates the initial out-of-plane orientation of $\vec{p}$ in a {Mn$_{3}$Sn} slab after applying a prior field of $-$60 mT along $[2\bar{1}\bar{1}0]$, which is evidenced by the in-plane temperature gradient in the measured $X_{\rm{odd}}$ image obtained at $J_{\rm{c}}$ = 50 mA along $[0001]$. Fig. 5b-5f illustrates the out-of-plane rotation of $\vec{p}$ during an in-plane field-scan cycle, along with the corresponding $X_{\rm{odd}}$ images. In Fig. 5b, 5d and 5f, uniform temperature distributions similar to that in Fig. 3a are found, indicating $\vec{p}$ alignment parallel to the in-plane field. In Fig 5c and 5e, in-plane temperature gradient similar to that in Fig. 5a is found, which means that the initial out-of-plane orientation of $\vec{p}$ can be recovered at a low field during the in-plane field sweeping. Therefore, $\vec{p}$ shows an anti-clockwise rotation from $B_{\rm ip}$ = $-$240 mT to $B_{\rm ip}$ = 240 mT and a clockwise rotation from $B_{\rm ip}$ = 240 mT to $B_{\rm ip}$ = $-$240 mT. Fig. 5g-5l shows the measured $X_{\rm{odd}}$ images for the in-plane field sweep after a reversed prior field, $B_{\rm op}$ = 60 mT. The thermal images show that $\vec{p}$ is aligned parallelly with the in-plane field at $B_{\rm ip}$ = $\pm$240 mT but has an opposite out-of-plane orientation in the low field range compared to those in Fig. 5c-5e, evidencing the prior-field-dependent rotation chirality of $\vec{p}$. In previous studies, the prior-field-dependent rotation chirality of $\vec{p}$ in the ``memory effect" was proposed to understand the planar AHE and planar ANE during an in-plane magnetic field sweeping\cite{Li2019,Xu2020Planar}, which was not directly observed. Here the LIT measurement clarifies the microscopic origin of the ``memory effect" in \MnSn. Compared to MOKE which allows observation of the $\vec{p}$ component along one direction in an optical setup\cite{Higo2018large,Uchimura2022}, our method offers a clear benefit for investigating 3D octupole moments dynamic processes.

\section{CONCLUSIONS}\label{sec3}
In summary, we have developed a new technique for visualizing the octupole domain structure in non-collinear AFMs using the AEE at room temperature. The AEE imaging of the domain configurations in non-collinear AFMs may well accelerate the development of future memory devices and facilitate the study of current-driven domain dynamics. Our preliminary \textit{ab initio} calculation shows the AEE is sizable in several materials in the {Mn$_{3}$X\xspace} family, indicating its wide application. Furthermore, some non-collinear \AFIC materials may exhibit a small Kerr angle but a large AEE coefficient, for which this technique is particularly advantageous.

\section{METHODS}\label{sec4}
\subsection{Preparation of \MnSn samples}
The \MnSn single crystals were grown by the flux method using Bi as the flux. Mn powder (99.95\% purity), Sn granules (99.999\% purity), and Bi granules (99.999\% purity) were mixed in a 3:1:3 molar ratio. This mixture was loaded into an alumina crucible, sealed in a quartz ampoule under a partial argon atmosphere, and heated to 1150 $^{\circ}$C in a furnace within 15 hours, maintaining this temperature for 10 hours. The assembly was then slowly cooled to 700 $^{\circ}$C at a rate of 50 $^{\circ}$C per hour, held at that temperature for 10 hours, and further gradually cooled to 300 $^{\circ}$C at a rate of 2 $^{\circ}$C per hour, maintaining it for 50 hours. Finally, the assembly was taken out from the furnace, and \MnSn crystals were separated from the flux using a centrifuge.
\subsection{Sample characterization} 
The LIT measurement was made by the Luxet thermo 100 system produced by Suzhou Luxet Infrared Technology Co., Ltd. The magnetization was measured using a commercial superconducting quantum interference device magnetometer (MPMS, Quantum Design). The Hall resistivity was measured using a standard four-probe geometry with a 10 mA current. Magneto-optical imaging of \MnSn crystals with $(2\bar{1}\bar{1}0)$ oriented surface plane was performed at room temperature using a Kerr microscope with 620 nm wavelength LED (TuoTuo Technology). 
\subsection{Experimental setup}
The LIT system is composed of an infrared camera, a microscope lens, a system processing unit that performs real-time Fourier analysis of detected thermal images, and a source meter. Electromagnets situated underneath the infrared camera enable the measurement of thermal images under in-plane or out-of-plane magnetic fields. To prevent vibrations from interfering with the measurements, the camera and electromagnet are situated on a vibration isolation table. All \MnSn samples used in this study were unpolished and their surfaces were coated with insulating black ink for LIT measurements.

\section{ACKNOWLEDGEMENTS}\label{sec6}
The authors thank Z.G. Liang and L.F. Wang for the magnetization measurement. The study was supported by the supercomputing service of USTC, and the USTC Center for Micro- and Nanoscale Research and Fabrication. 

\section{FUNDING}\label{sec7}
This work was supported by the National Key R\&D Program (2022YFA1403502), the National Natural Science Foundation of China (12234017,12074366), the  Fundamental Research Funds for the Central Universities (WK9990000116). Y.F.G. acknowledges the support from the Shanghai Science and Technology Innovation Action Plan (21JC1402000) and the Double First-Class Initiative Fund of ShanghaiTech University. Y. Gao is supported by the  Fundamental Research Funds for the Central Universities (WK2340000102). Z. Liu is supported by the National Natural Science Foundation of China (11974327 and 12004369), Fundamental Research Funds for the Central Universities (WK3510000010, WK2030020032), Anhui Initiative in Quantum Information Technologies (AHY170000), and Innovation Program for Quantum Science and Technology (2021ZD0302800).

\section{AUTHOR CONTRIBUTIONS}\label{sec8}
D.H., Y.F.G., P.W., and W.X. designed the experiment, W.X., Z.X., P.W., and J.S. fabricated the samples and collected all of the data. Y.C., W.P., J.Z., H.P., X.Y. and H.Y. contributed to the experimental set-up. P.W., D.H., and Y.F.G. analyzed the data. Q.N., Y.G. and Z.L. contribute theoretical discussions. D.H. and Y.F.G. supervised this study. All the authors discussed the results and prepared the manuscript.
~\\\\
\textbf{Conflict of interest statement.} None declared.


\end{document}


\title{Supplementary Information for ``Infrared Imaging of Magnetic Octupole Domains in Non-collinear Antiferromagnets''}


\maketitle

\tableofcontents

\subsection{MOKE images of \MnSn}
Figure S1 shows the MOKE images of \MnSn. Kerr signal can not be detected probably due to surface oxidization and roughness. MOKE images of a 0.7-nm-thick Co film with perpendicular magnetic anisotropy (PMA) under the same set-up is also shown for comparison.

\renewcommand*{\thefigure}{\textbf{S\arabic{figure}}}
\renewcommand{\figurename}{\textbf{Supplementary Fig.}}
\begin{figure}[h!]
	\centering
	\includegraphics[width=0.6\textwidth]{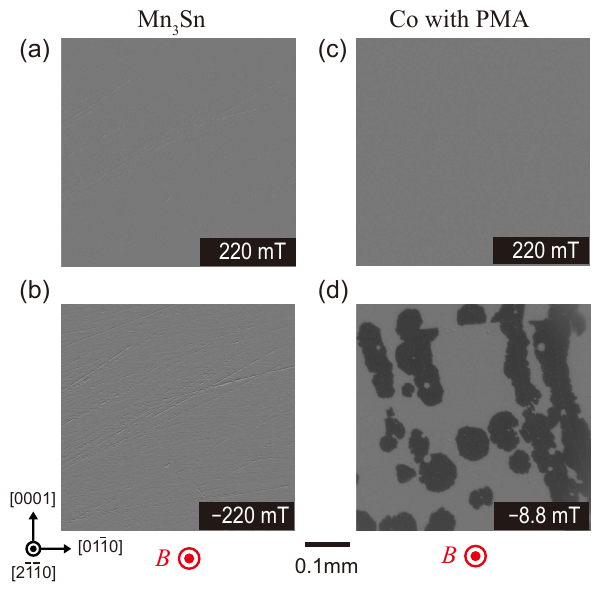}
	\caption{ \textbf{MOKE images of \MnSn.} 
\textbf{a},\textbf{b}, MOKE images of \MnSn obtained at an out-of-plane field $B$ = 220 mT (\textbf{a}) and $B$ = $-$220 mT (\textbf{b}). \textbf{c},\textbf{d}, Moke images of a 0.7-nm-thick Co film with perpendicular magnetic anisotropy (PMA) for comparison obtained at $B$ = 220 mT (\textbf{c}) and $B$ = $-$8.8 mT (\textbf{d}). Grey and black regions correspond to positive and negative values of the MOKE signal.}
	\label{figS1}
\end{figure}

\subsection{The AHE and the optical image of the \MnSn sample in Fig. 2}
Figure S2(a) shows the measured anomalous Hall effect (AHE) signal of the same \MnSn sample in Fig. 1 and Fig. 2 in the main text. Figure S2(b) shows the optical image of this sample, which is covered with insulating black ink for lock-in thermography measurements.

\begin{figure}[h!]
	\centering
	\includegraphics[width=0.7\textwidth]{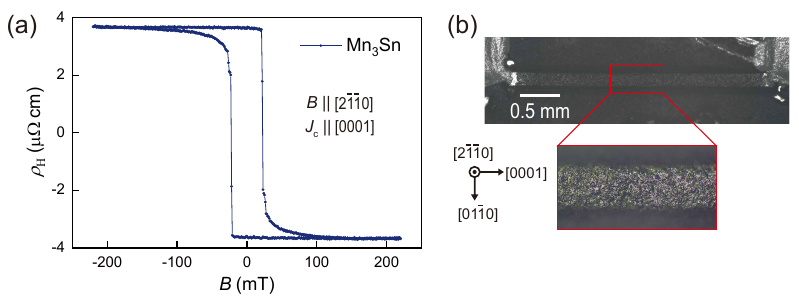}
	\caption{ \textbf{The AHE (a) and the optical image (b) of the \MnSn sample in Fig. 2.}}
	\label{figS2}
\end{figure}

\subsection{Compare the $X$ curves to magnetization $M$}

\begin{figure}[h!]
	\centering
	\includegraphics[width=0.8\textwidth]{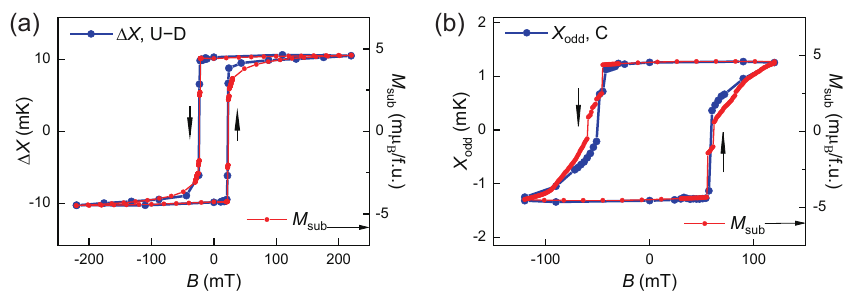}
	\caption{ \textbf{Compare the ${\Delta}X$ (or $X_{\rm{odd}}$) curves in Fig. 2 and Fig.  3 to magnetization $M$ subtracted by a linear function.}}
	\label{figS3}
\end{figure}

\subsection{The out-of-plane AEE results of the same \MnSn sample in Fig. 4}
Figure S4 shows the $X_{\rm{odd}}$ images and the corresponding ${\Delta}X_{\rm{odd}}$ curve obtained in the out-of-plane field-scan cycle of the same \MnSn sample in Fig. 4 in the main text.

\begin{figure}[h!]
	\centering
	\includegraphics[width=0.8\textwidth]{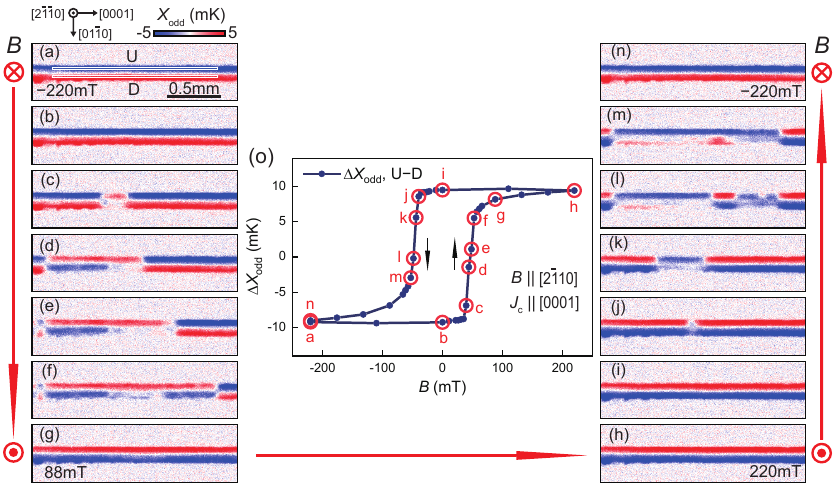}
	\caption{ \textbf{The out-of-plane AEE results of the same \MnSn sample in Fig. 4.} 
\textbf{a}-\textbf{n}, $X_{\rm{odd}}$ images obtained in the out-of-plane field-scan cycle from $-$220 mT to 220 mT along $[2\bar{1}\bar{1}0]$ at $J_{\rm{c}}$ = 50 mA along $[0001]$, where $X_{\rm{odd}}$ denotes the lock-in temperature modulation with the $B$-odd dependence. \textbf{o}, Field dependence of ${\Delta}X_{\rm{odd}}$ calculated from the U and D areas of the \MnSn slab, where the plotted data were obtained by subtracting the average values on D from U.}
	\label{figS4}
\end{figure}

\subsection{Movie of the out-of-plane magnetic reversal process in \MnSn}
\textbf{Supplementary Movie S1.} The movie of the out-of-plane magnetic reversal process under a more precise magnetic field scanning with a step of 0.01 mT for the same \MnSn sample in Fig. 2 in the main text.

\subsection{The definition of the anomalous Ettingshausen coefficient}
There are two prevalent definitions for the anomalous Ettingshausen effect (AEE) coefficient in published literatures. The first defines the AEE coefficient as $\varepsilon_{\rm{AEE}}=\nabla_{y}T/j_{x}$\cite{Behnia2015book,Xu2020-AEE}, where $\nabla_{y}T$ is the transverse thermal gradient generated by the AEE, and $j_{x}$ is the longitudinal current density. This definition is equivalent to the one used in our study, $\nabla T_{\rm{AEE}} = \varepsilon_{\rm{AEE}} (\vec{j}_{\rm{c}}\times\vec{p})$. Under this definition, the Bridgman relation, linking the AEE coefficient $\varepsilon_{\rm{AEE}}$ and the anomalous Nernst effect (ANE) coefficient $S_{\rm{ANE}}$, is expressed as $\varepsilon_{\rm{AEE}}=S_{\rm{ANE}}T/\kappa$\cite{Bridgman1924,Callen1948}, with $T$ being the absolute temperature and $\kappa$ being the thermal conductivity. 

The second definition characterizes the AEE coefficient as $\vec{j}_{\rm{q,AEE}} = \Pi_{\rm{AEE}} (\vec{j}_{\rm{c}}\times\vec{m})$\cite{Seki2018,Miura2019}, where $\vec{j}_{\rm{q,AEE}}$, $\Pi_{\rm{AEE}}$, $\vec{j}_{\rm{c}}$, and $\vec{m}$ denote the heat current density generated by the AEE, the AEE coefficient, the charge current density, and the unit vector of magnetization $\vec{M}$, respectively. Under this definition, the Brigman relation is $\Pi_{\rm{AEE}}=S_{\rm{ANE}}T$, which exhibits a formal similarity to the Kelvin relation $\Pi=ST$, with $\Pi$ being the Peltier coefficient and $S$ being the Seebeck coefficient. 

Both definitions of the AEE coefficient are commonly employed, differing only in terms of the thermal conductivity $\kappa$. Since we did not directly measure thermal conductivity, we have adopted the first definition in our work.